\documentclass[referee, pdflatex, sn-nature]{sn-jnl-modified}

\usepackage{graphicx}
\usepackage{multirow}
\usepackage{amsmath,amssymb,amsfonts}
\usepackage{amsthm}
\usepackage{mathrsfs}
\usepackage[title]{appendix}
\usepackage{xcolor}
\usepackage{textcomp}
\usepackage{manyfoot}
\usepackage{booktabs}
\usepackage{algorithm}
\usepackage{algorithmicx}
\usepackage{algpseudocode}
\usepackage{listings}

\usepackage{bm}
\usepackage{lineno}

\begin{document}

\title[Article Title]{Observation of Superfluidity and Meissner Effect of Composite
Bosons in GaAs Quantum Hall System}

\author[1,5]{\fnm{Yuanze} \sur{Li}}
\author[2,5]{\fnm{Renfei} \sur{Wang}}
\author[1,5]{\fnm{Jiahao} \sur{Chen}}
\author[2]{\fnm{Wenfeng} \sur{Zhang}}
\author[3]{\fnm{Adbhut} \sur{Gupta}}
\author[3]{\fnm{Kirk W.} \sur{Baldwin}}
\author[3]{\fnm{Loren} \sur{Pfeiffer}}
\author[2]{\fnm{Rui-Rui} \sur{Du}}
\author*[2]{\fnm{Yang} \sur{Liu}}\email{liuyang02@pku.edu.cn}
\author*[1,4]{\fnm{Tian} \sur{Liang}}\email{tliang@mail.tsinghua.edu.cn}

\affil[1]{\orgname{State Key Laboratory of Low Dimensional Quantum Physics, Department of Physics, Tsinghua University}, \orgaddress{\city{Beijing} \postcode{100084}, \country{People's Republic of China}}}

\affil[2]{\orgname{International Center for Quantum Materials, School of Physics, Peking University}, \orgaddress{\city{Beijing} \postcode{100871}, \country{People's Republic of China}}}

\affil[3]{\orgname{Department of Electrical Engineering, Princeton University}, \orgaddress{\city{Princeton}, \state{New Jersey} \postcode{08544}, \country{USA}}}

\affil[4]{\orgname{Frontier Science Center for Quantum Information}, \orgaddress{\city{Beijing} \postcode{100084}, \country{People's Republic of China}}}

\affil[5]{These authors contributed equally to this work}

\abstract{

The quantum Hall effect (QHE) is theoretically understood as a superfluid condensate of composite bosons (CBs)---bound states of electrons and magnetic flux quanta. While dissipationless transport is consistent with this picture, other signatures of superfluidity, such as the Meissner effect, remain elusive. Here, we present direct experimental evidence for CB superfluidity by probing the system's response to a controlled, time-varying magnetic field in Corbino disk geometries. We simultaneously observe the quantized Laughlin charge pumping and a new, quantized charge accumulation phenomenon, governed by the relation $\Delta Q_{\rm a}/e = \nu\,(\Delta \Phi/\Phi_0)$. This relation signifies that the system actively maintains the fixed electron-to-flux ratio that defines the CBs, neutralizing excess flux by drawing in a precise number of electrons.

Crucially, devices with multiple concentric top gates reveal that this charge accumulation is uniformly distributed across the bulk of the QHE fluid, demonstrating that it is a collective, bulk property rather than an edge effect---a key signature of a superfluid condensate. Furthermore, the presence of a top gate determines the screening mechanism: in a "grand canonical" setting with a gate, low Coulomb energy favors a charge-mediated screening (generalized Meissner effect); without a gate, the system enters a "canonical" regime, exhibiting fixed electron density like type-II superconductors. These observations confirm the CB superfluid nature of the QHE ground state and establish a versatile platform for studying macroscopic quantum coherence and its screening transitions in two dimensions.
}

\maketitle

\section{Introduction}

Since the discovery of the quantum Hall effect (QHE) \cite{klitzing1980, tsui1982FQHE}, its microscopic theory has remained a central and fruitful field of research for decades. The Laughlin wave function successfully explains the strongly correlated bulk states of the fractional quantum Hall effect (FQHE) \cite{laughlin1983wavefunc}, including the $\nu = 1$ state. However, an intuitive physical picture for the QHE in GaAs was not fully elucidated until the composite boson (CB) picture was introduced, explaining the QHE state as a superfluid condensate \cite{girvin1987, read1989, zhang1992, haldane2011}. In this theory, QHE states at filling factors $\nu = p/q$ are described as CBs, each consisting of $p$ electrons bound to $q$ flux quanta, acquiring a total statistical exchange phase of $(-1)^{pq-p} = +1$ for both the integer (IQHE) and fractional (FQHE) effects \cite{haldane2011}. When the GaAs QHE system forms CBs, it exhibits off-diagonal long-range order (ODLRO), allowing macroscopic propagation without friction—the hallmark of superfluidity \cite{gorkov1958, yang1962ODLRO, girvin1987}. In such a state, a supercurrent of CBs flowing along the $x$-axis automatically generates a transverse Hall electric field along the $y$-axis due to the movement of flux quanta, given by $\bm{E} = n_\Phi \Phi_0\bm{e}_z \times \bm{v} = -h/(\nu e^2) \bm{e}_z \times \bm{j}$, where $n_\Phi$ is the flux quantum density, $\Phi_0 = h/e$ is the flux quantum, and $\bm{v}$ is the group velocity of CBs, as required by Lorentz covariance.

The vanishing longitudinal resistivity in the QHE is consistent with the dissipationless nature of superfluidity. However, other critical signatures of CB superfluidity—such as the Meissner effect—remain largely unverified experimentally. In the superfluid theory of QHE ground states, all external magnetic field is effectively incorporated into the CBs, leaving an effective zero field, consistent with the Meissner effect. Therefore, measuring the electromagnetic response of the QHE system to external magnetic field variations, analogous to experimental methodologies in superfluidity, may reveal crucial information about its superfluid nature. Importantly, the Laughlin charge pumping model describes such a field-variation-induced phenomenon \cite{laughlin1981}, where electrons are transferred through the QHE bulk driven by a gauge variation induced by a changing field outside the system. This model has been experimentally implemented to measure the 2D bulk Hall conductance, first achieved in a Corbino-disk-shaped GaAs sample under a sweeping DC magnetic field \cite{1992_Russia_GaAs_chargepumping_and_Streda}. Notably, a difference in transferred charge between the inner and outer electrodes was detected, indicating charge accumulation inside the sample. However, neither a sound theoretical explanation nor accurate experimental quantization was achieved. Subsequent experiments measured Laughlin charge pumping in GaAs QHE \cite{1995_GaAs_chargepumping} systems via AC field methods, but the charge accumulation phenomenon was not investigated.

We propose that the observed charge accumulation is generated directly by magnetic field variations penetrating the QHE system, providing distinctive evidence for CB superfluidity (Fig. \ref{fig:fig1}(a)). From a simplified, localized perspective, the charge accumulation can be explained by the Hall current of a circulating Faraday electric field generated by $\Delta B$, leading to an accumulated charge density of $\eta = \sigma_{xy} \Delta B = \frac{\nu e^2}{h} \Delta B$ (where $\sigma_{xy}$ is the Hall conductance; see Supplementary Materials). The superfluid picture offers a natural and profound microscopic understanding: the superfluid rigidity tends to maintain the CB condensate ground state without extra magnetic flux, as extra flux would excite quasiparticles and cost energy. The charge accumulation is a precise response of the electron density to the flux quantum density, effectively eliminating extra flux through CB formation and resulting in a fixed ratio of electrons to flux quanta\footnote{This fixed ratio of electrons to flux quanta is conventionally called incompressibility.}. Since the ratio between electrons and flux quanta in a CB equals the filling factor $\nu$, the expected charge accumulation satisfies $Q_\text{a} / e = \nu (\Delta\Phi / \Phi_0)$, where $Q_\text{a} = \eta A$ is the total accumulated charge ($A$ is the sample area) and $\Delta\Phi$ is the extra flux. For a uniform field variation, an evenly distributed charge accumulation density $\eta = \frac{\nu e^2}{h} \Delta B$ is anticipated. This scenario requires that the energy gain from CB superfluidity outweighs the Coulomb potential loss, effectively placing the system in a grand canonical ensemble that allows electron density change. As we will show, this corresponds to the case for samples with a grounded top gate. In contrast, without a top gate, the system is effectively in a canonical ensemble with a strictly fixed electron density. In this case, CB superfluidity still holds and manifests like type-II superconductors, with fixed density. Our subsequent measurements verify these theoretical predictions step by step.

It is important to recognize that the original Laughlin charge pumping model does not include charge accumulation because the magnetic field inside the QHE region is assumed unchanged. However, in actual experiments, magnetic field variations occur both outside and within the sample area, causing charge pumping and charge accumulation to coexist. Meanwhile, the microscopic CB theory coherently explains the quantized charge pumping (see below for details). In the following, we first confirm our experimental technique by simultaneously measuring charge accumulation and Laughlin charge pumping to obtain quantized results, before discussing the experimental manifestation of the CB superfluidity nature in the QHE system.

\section{Experiment}

\subsection{Experimental setup for charge detection}
\label{section2.1}

\begin{figure}[htbp]
  \centering
  \includegraphics[width=1\linewidth]{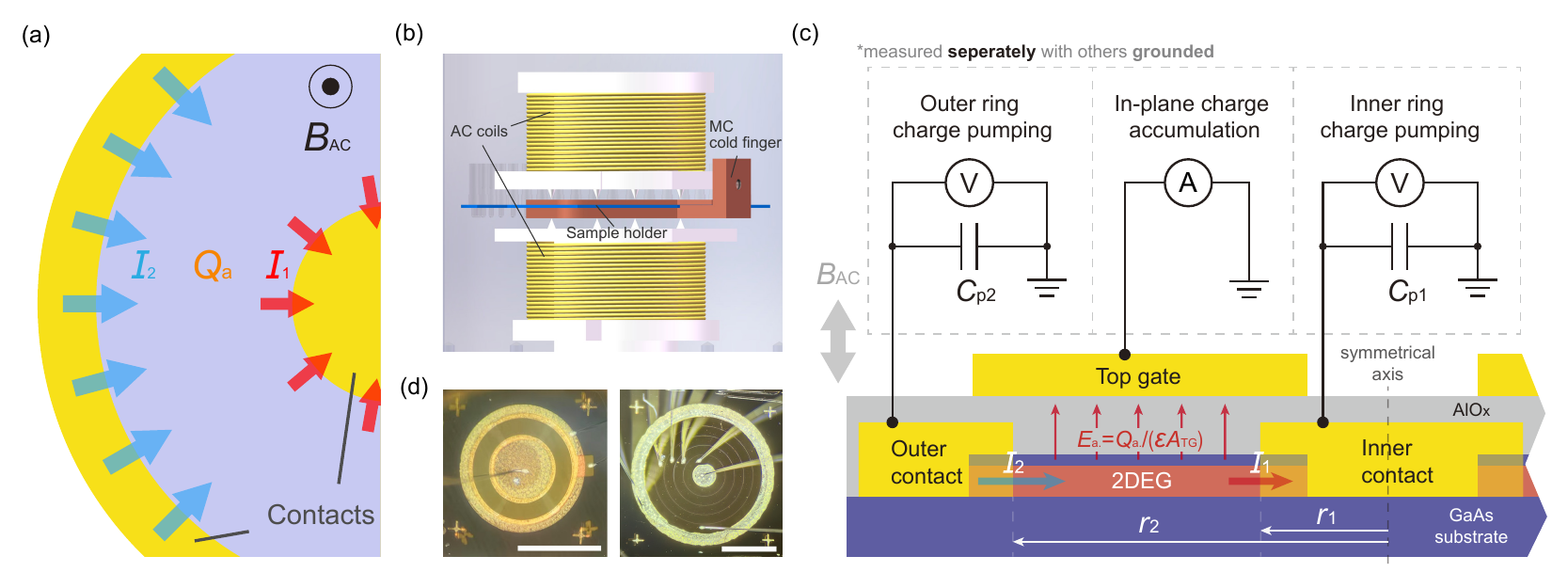}
  \caption{\textbf{Experimental setup and sample structure.} (a) The schematic of the charge pumping ($I_1$ and $I_2$) and charge accumulation ($Q_\text{a}$) under uniform $B_\text{AC}$. $B_\text{AC}$ penetrating the sample area results in $Q_\text{a}$ and a difference between $I_1$ and $I_2$ (all AC signals). (b) The AC field generation setup. Samples were placed inside the AC coil structure and cooled by the mixing chamber (MC). (c) Sample structures and measurement circuits of a Corbino-disk-shaped sample. $I_1$ and $I_2$ were measured from the voltage across capacitances $C_\text{p1}$ and $C_\text{p2}$, while $Q_\text{a}$ was measured by a current preamplifier. Each of the three circuits was measured separately while the others were connected to ground. For samples with multiple top gates, all $Q_\text{a}$ were measured simultaneously. Coulomb potential rise of accumulated electrons is greatly reduced by the capacitance coupling of the top gate, determined by the thickness of the dielectric layer and $E_\text{a}$ generated by $Q_\text{a}$. (d) Photos of single-gate and multi-gate samples (Devices I and IV).}
  \label{fig:fig1}
\end{figure}

We use Corbino-disk-shaped samples (annular with inner and outer contact electrodes) fabricated from high-mobility GaAs/AlGaAs quantum wells to generate charge pumping and charge accumulation signals. These samples have top gates covering the QHE region, which are crucial for detecting the accumulated charge. When the top gates are grounded, the Coulomb potential energy generated by the charge accumulation ($Q_\text{a}$) is greatly reduced by capacitive coupling, stabilizing the electrons in the QHE state (see below for details). Moreover, the capacitive coupling induces a charge under the top gate opposite to the accumulated charge, while the charge flowing out of the gate equals the accumulated charge. Therefore, the top gates function as charge accumulation detectors.

Fig. \ref{fig:fig1}(b) shows a schematic of our field generation setup. A coil system inside a dilution refrigerator generates a controllable AC magnetic field ($B_\text{AC}$) at low temperature (below 100 mK), serving as the source of flux variation. This AC field is perpendicular to the sample plane and parallel to the DC field ($B_\text{DC}$) hosting the QHE.

Laughlin charge pumping and charge accumulation signals are measured by different circuits for optimal signal-to-noise ratio, as illustrated in Fig. \ref{fig:fig1}(c). For charge pumping, the charge from contacts is converted into voltage signals $U_\text{pump} = Q_\text{pump}/C_\text{p}$ via parallel capacitances $C_\text{p1}$ and $C_\text{p2}$. For charge accumulation, the charge leaving the top gate is detected by current preamplifiers held at ground potential. To avoid mutual interference, the signals presented here were measured separately, with the other interfaces grounded. (Simultaneous measurements yield nearly identical results; see Supplementary Materials for details.)

Photos of real samples (Devices I and IV) are shown in Fig. \ref{fig:fig1}(d). Single-gate samples are measured to determine the charge accumulation value (Sections \ref{section2.2} and \ref{section2.4}), while samples with segmented multiple top gates provide spatial information on charge distribution within the QHE bulk, as each gate only detects charge within its region (Section \ref{section2.3}). This is essential for corroborating the anticipated uniform distribution arising from superfluidity in a grand canonical ensemble.

\subsection{Quantized charge accumulation and charge pumping}
\label{section2.2}

\begin{figure}[htbp]
  \centering
  \includegraphics[width=1\linewidth]{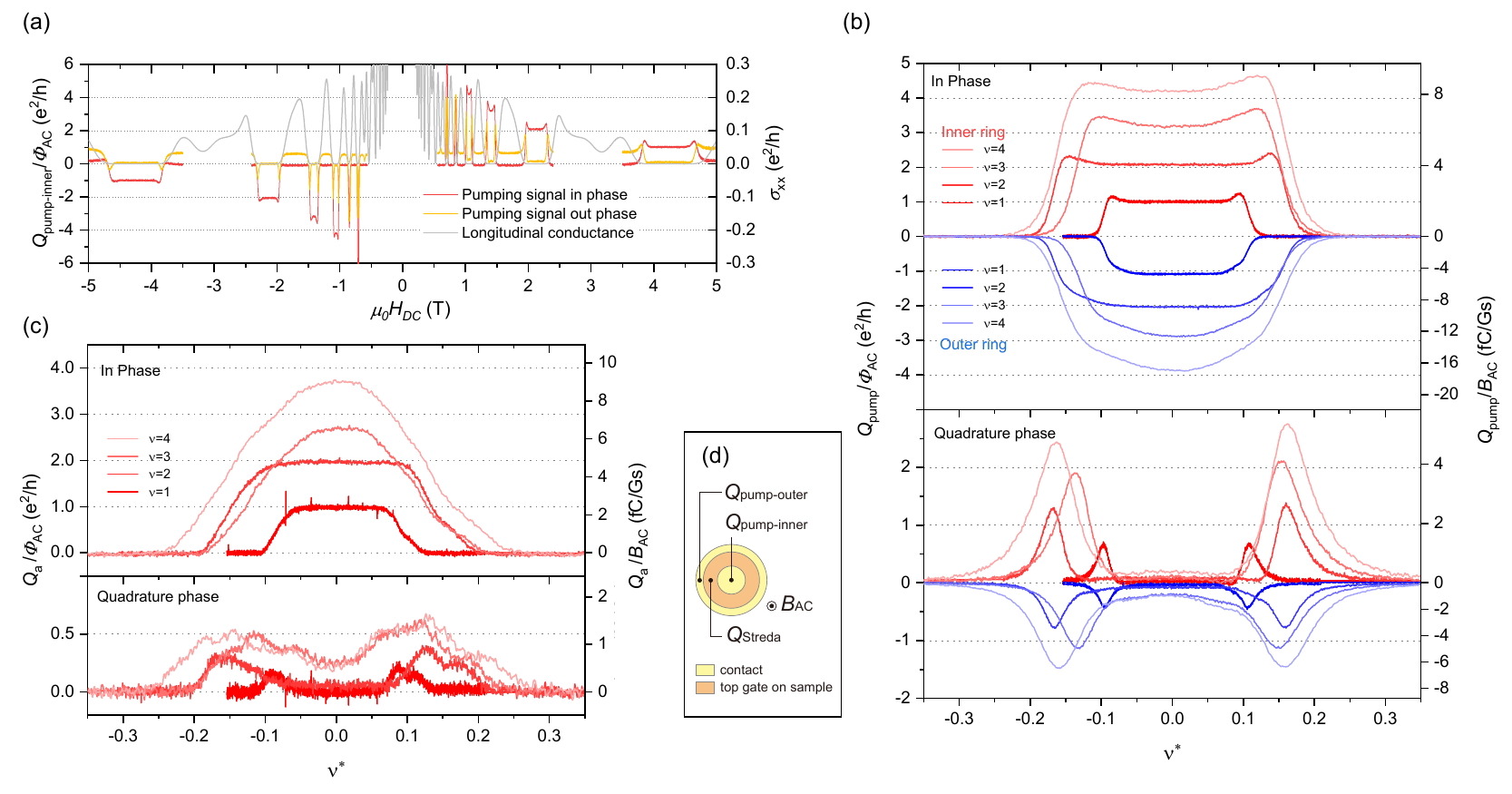}
  \caption{\textbf{Quantized charge pumping and charge accumulation in single-top-gate devices.} $Q_\text{pump-inner}$, $Q_\text{pump-outer}$, and $Q_\text{a}$ signals were measured from a Corbino-disk-shaped sample with contact radii of 400 to 600 µm (Device I). Each charge signal is divided by the effective flux variation $\Phi_\text{AC}$ respectively ($\pi r_1^2 B_\text{AC}$ for inner charge pumping, $\pi r_2^2 B_\text{AC}$ for outer charge pumping, and $\pi (r_2^2-r_1^2) B_\text{AC}$ for charge accumulation) to show the quantization. 
  (a) The raw signals of $Q_\text{pump-inner}$ are compared with the $\sigma_{xx}$ curve of the same sample. Quantized in-phase plateaus of $Q_\text{pump-inner}/\Phi_\text{AC}$ appear, consistent with vanishing $\sigma_{xx}$. 
  (b) Antisymmetrized (same as below) $Q_\text{pump-inner}$ and $Q_\text{pump-outer}$ are plotted against the relative filling factor $\nu^* = B_{(\nu = 1)}/B_\text{DC}-\nu$ ($B_{(\nu = 1)} = 4.23$ T). In-phase $Q_\text{pump}/\Phi_\text{AC}$ signals show clear quantized plateaus, while quadrature signals vanish to zero within the plateaus, revealing the consistent phase between $Q_\text{pump}$ and $B_\text{AC}$. The right vertical axis $Q_\text{pump}/B_\text{AC}$ has different scales at upper and lower quadrants, showing the difference between $Q_\text{pump-inner}$ and $Q_\text{pump-outer}$. 
  (c) $Q_\text{a}$ signals plotted against $\nu^*$. The same vertical axes are used as in (b). In-phase components of $Q_\text{a}/\Phi_\text{AC}$ also show quantized plateaus, with quadrature components vanishing.
  (d) Sketch of charge measurement on single-gate samples. Measurement circuits are described in Fig. \ref{fig:fig1}(c).
  }
  \label{fig:fig2}
\end{figure}

Charge pumping signals from a single-gate Corbino-disk-shaped sample (Device I) are shown in Fig. \ref{fig:fig2}(a). The sample has inner and outer ring radii of $r_1 = 400$ µm and $r_2 = 600$ µm, respectively. The pumped charge at the inner ring ($Q_\text{pump-inner}$) is compared with the longitudinal conductance $\sigma_{xx}$, measured via a two-terminal method between two contacts. For quantitative analysis, the vertical axis shows $Q_\text{pump-inner}$ divided by the effective AC magnetic flux $\Phi_\text{AC} = \pi r_1^2 B_\text{AC}$, where $\pi r_1^2$ is the area enclosed by the inner ring. At each integer filling factor $\nu$ ($\nu \leq 4$), the in-phase signals exhibit quantized plateaus at $\nu e^2/h$, corresponding to the Hall conductance $\sigma_{xy}$. The plateau regions align precisely with those of $\sigma_{xx}$, consistent with charge pumping theory. Outside the plateaus, an out-of-phase background signal becomes significant, primarily arising from the uncanceled Faraday effect and crosstalk from AC coil excitation. Antisymmetrization with respect to the DC magnetic field effectively removes this background and manifests the consistent phase between $Q_\text{pump}$ and $B_\text{AC}$ (see Supplementary Materials for details). All signals discussed and plotted hereafter are antisymmetrized unless specified.

Our theory suggests that field variation within the QHE region causes charge accumulation and differing pumped charges at the two edges, unlike the original model which predicts identical charges. To investigate this, we present charge pumping signals ($Q_\text{pump}$) measured at both inner and outer rings (Fig. \ref{fig:fig2}(b)). The positive sign is defined as charge flowing out of the sample. The horizontal axis uses the relative filling factor $\nu^* = B_{(\nu = 1)}/B_\text{DC}-\nu$ for clarity (data from the $\nu=1,2,3,4$ plateaus are shown; $B_{(\nu = 1)}$ is the magnetic field at exact $\nu = 1$). The right vertical axis shows $Q_\text{pump}$ normalized by $B_\text{AC}$, while the left axis shows $Q_\text{pump}$ normalized by the AC magnetic flux enclosed by the corresponding sample edge ($\Phi_\text{AC} = \pi r^2 B_\text{AC}$). The in-phase signals at both rings consistently show quantized plateaus agreeing with $\sigma_{xy} = \nu e^2/h$, in line with Laughlin's theory. Crucially, while $Q_\text{pump}/\Phi_\text{AC}$ signals are quantized at both rings, the absolute charge values ($Q_\text{pump}$) differ, with a ratio consistent with $r_1^2/r_2^2 = 4:9$ between inner and outer rings (see right axis). This difference reveals a charge accumulation within the sample area, which can be quantitatively characterized from interior flux variations. The out-of-phase signals show double-peaked structures accompanying the decay of the in-phase quantized plateaus, dominated by the surge of $\sigma_{xx}$ upon QHE breakdown. $Q_\text{pump}$ signals are stable against frequency and temperature (below 1 K) variations near exact filling for lower filling factors ($\nu =1,2$). Near plateau edges or at higher fillings ($\nu = 3,4$), signals exhibit temperature and frequency dependence due to increasing $\sigma_{xx}$, similar to the behavior at plateau edges (see Supplementary Materials for details).

Using the same single-gate sample, we measured charge accumulation signals ($Q_\text{a}$) via the top gate (Fig. \ref{fig:fig2}(c)), retaining the relative filling factor $\nu^*$ and dual vertical axes ($Q_\text{a}/\Phi_\text{AC}$ and $Q_\text{a}/B_\text{AC}$) as in Fig. \ref{fig:fig2}(b). Here, $\Phi_\text{AC} = \pi (r_2^2-r_1^2) B_\text{AC}$ is defined using the total sample area $\pi (r_2^2-r_1^2)$. Charge accumulation appears and exhibits signal plateaus within the quantized region, where the in-phase $Q_\text{a}/\Phi_\text{AC}$ signals show quantization at $\nu e^2/h$, consistent with the charge-per-flux ratio of CBs. Signals for larger filling factors ($\nu =3,4$) slightly deviate from quantization due to larger $\sigma_{xx}$, consistent with the decay of charge pumping. We also performed measurements on single-gate samples with different geometries, yielding results consistent with theory (see Supplementary Materials for details). Specifically, for single-gate samples, the charge accumulation can be calculated from the pumped charge due to charge conservation (see Supplementary Materials for details). In summary, the quantization of both charge pumping and charge accumulation in single-gate samples is quantitatively measured, supporting both experimental validity and the superfluid theoretical model.

\subsection{Superfluidity and uniform distribution of charge accumulation}
\label{section2.3}

\begin{figure}[htbp]
  \centering
  \includegraphics[width=1\linewidth]{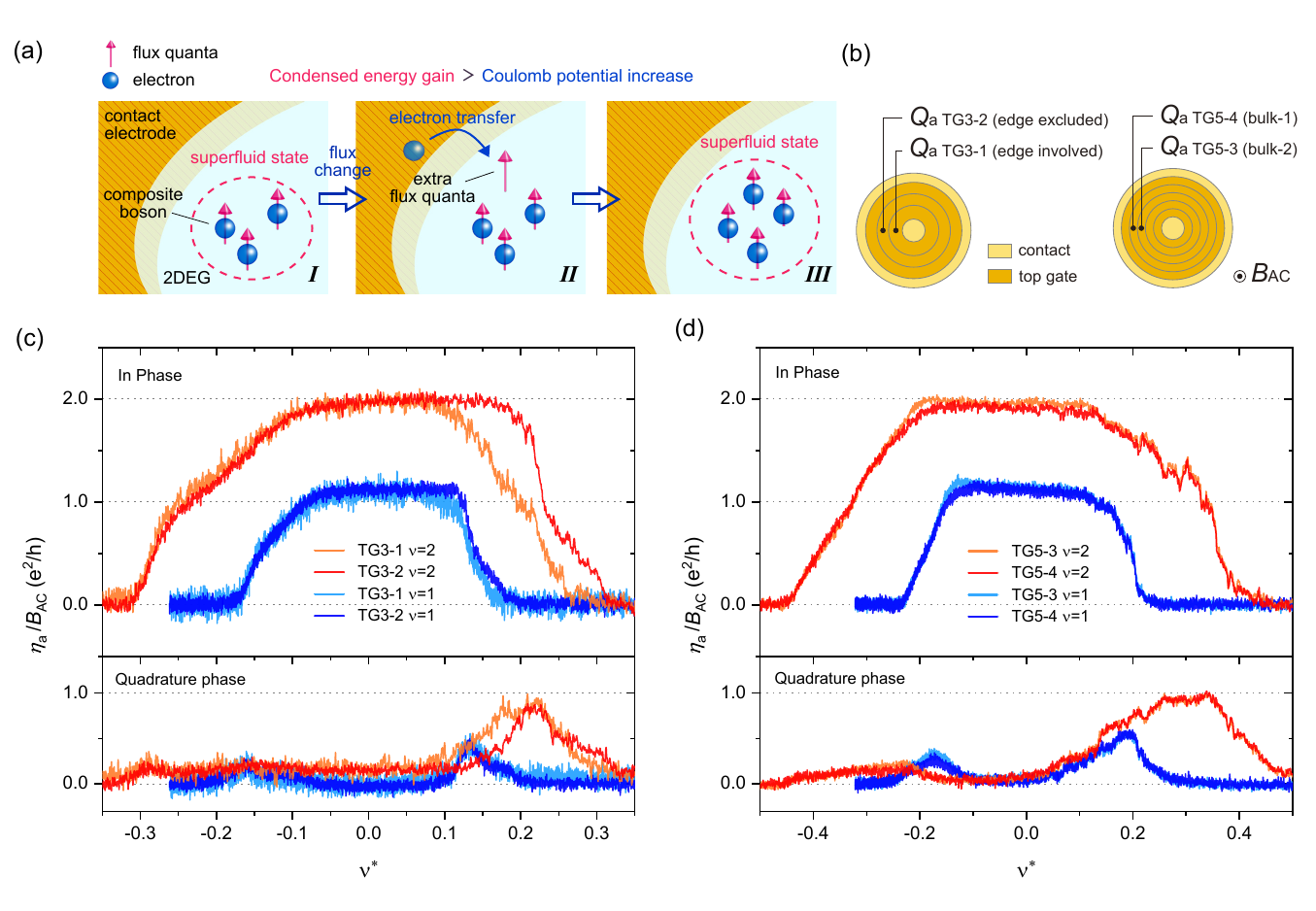}
  \caption{\textbf{Condensate of CBs and uniformly distributed charge accumulation confirmed by multi-top-gate samples.} 
  (a) Schematic of CB superfluid at $\nu=1$ and the dynamic charge accumulation process responding to flux variations. The contact of the sample places the system in a grand canonical ensemble for electrons, while the gate of the sample greatly reduces the Coulomb potential increase caused by accumulated charge. As a result, the transfer of electrons effectively absorbs extra flux quanta and preserves the condensed CB state to maintain the lowest energy of the grand canonical ensemble.
  (b) Sketch of charge measurement on multi-gate samples. $Q_\text{a}$ measured on different top gates is labeled. $Q_\text{a TG3-1}$ and $Q_\text{a TG3-2}$ compare the edge and bulk regions of the QHE system (Device III). $Q_\text{a TG5-3}$ and $Q_\text{a TG5-4}$ compare different bulk regions (Device IV).
  (c-d) $\eta_\text{a}/B_\text{AC}$ signals measured on multi-gate samples, where $\eta_\text{a}$ is the charge accumulation density calculated by $\eta_\text{a} = Q_\text{a}/A$ ($A$ the top gate area for each). Signals of $\nu=1$ and $\nu=2$ plateaus are plotted for every labeled gate, exhibiting consistent and quantized values.
  }
  \label{fig:fig3}
\end{figure}

Having established the existence of quantized charge accumulation, we now explore its intrinsic origin from the perspective of CB superfluidity. The experimentally observed relation, $\Delta Q_\text{a} / e = \nu (\Delta\Phi / \Phi_0)$, signifies that the system actively maintains a fixed ratio of electrons to flux quanta. Since the filling factor $\nu$ is precisely this ratio for a CB, our result indicates that the system remains in a pure CB state, free from residual electrons or flux quanta.

In other words, all introduced flux quanta are bound into newly formed CBs through electron transfer. This demonstrates that the QHE ground state is a robust condensate exhibiting superfluid rigidity against external magnetic field variations.

Quantitatively comparing energy scales, the electrostatic energy required to draw in charge is $E=Q^2/2C_\text{TG}$ (where $C_\text{TG}$ is the capacitance between the sample and top gate). Under our experimental conditions, the energy cost per electron is approximately $2 \times 10^{-4}$ meV ($\nu = 1$, $B_\text{AC} \sim 0.3$ Gs), much lower than the predicted superfluid energy gap (comparable to the cyclotron energy, at least on the order of $1 \times 10^{-1}$ meV in real QHE systems \cite{zhang1992}). A schematic of the dynamic charge accumulation process is shown in Fig. \ref{fig:fig3}(a). For clarity, the simplest case $\nu = 1$ is considered, where a CB consists of one electron and one flux quantum. An external field change introduces excess flux quanta that threaten to disrupt the condensate's coherent phase. To preserve superfluidity, the system neutralizes this perturbation by drawing in a precise number of electrons, which bind with the excess flux to form new CBs that seamlessly integrate into the condensate. This is a charge-mediated screening of the external field that preserves QHE superfluidity—a generalized Meissner effect.

This model predicts that screening should be a collective, bulk phenomenon. To test this crucial hypothesis of spatial uniformity, we fabricated multi-top-gate devices to spatially resolve the charge accumulation. The standard single top gate was segmented into a series of concentric, electrically isolated rings, allowing independent probing of different regions of the quantum Hall fluid. As shown in Fig. \ref{fig:fig3}(b), a three-gate device (Device III) compared the response between edge and bulk, while a five-gate device (Device IV) probed uniformity deep within the sample. During measurements, a calibrated AC magnetic field of 83.777 Hz was applied to induce charge accumulation, while target top gates were connected to current preamplifiers for simultaneous signal acquisition. All other gates and contact electrodes were held at ground.

Results for the $\nu=1$ and $\nu=2$ plateaus are presented in Fig. \ref{fig:fig3}(c) and (d). For every gate, regardless of location, the accumulated charge $Q_\text{a}$ is perfectly proportional to the magnetic flux passing through its area, following the quantized relation $Q_\text{a}/\Phi = \nu (e/\Phi_0)$. This confirms that charge accumulation is not a localized edge phenomenon but a uniform, collective response of the entire two-dimensional electron system. Such homogeneity is a key signature of a superfluid condensate: as the external field generates additional flux quanta uniformly across the sample, the system responds by uniformly drawing in electrons to form new CBs, preserving the condensed state throughout.

\subsection{Gate control of charge accumulation and preclusion of band-structure QHE}
\label{section2.4}

\begin{figure}[htbp]
  \centering
  \includegraphics[width=1\linewidth]{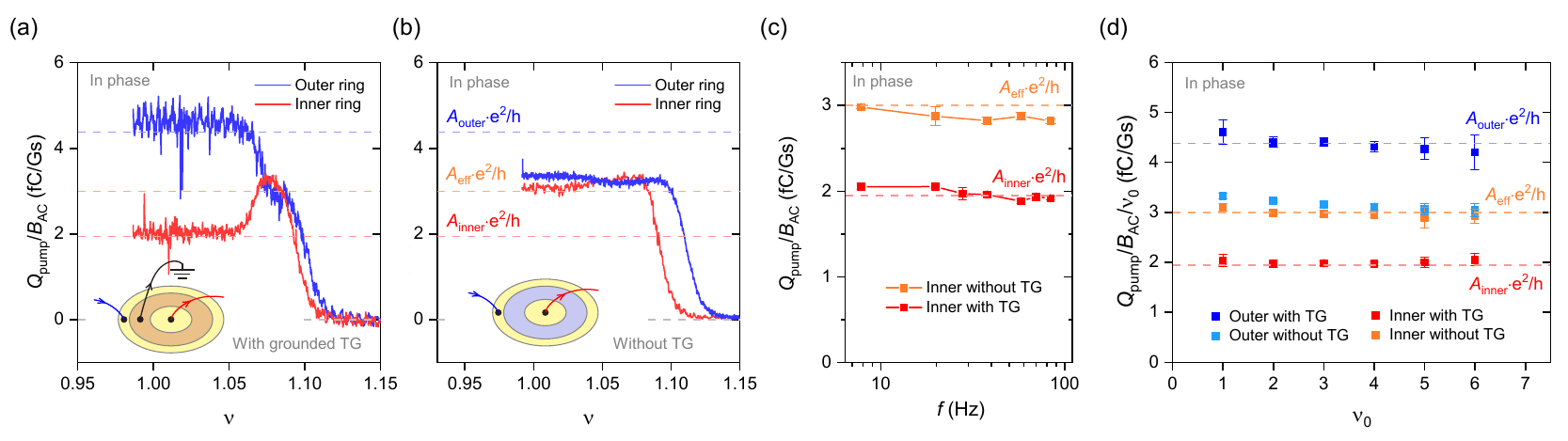}
  \caption{\textbf{Charge pumping signals controlled by the existence of the gate, reflecting the existence of charge accumulation.} Charge pumping signals of an ultrahigh mobility Corbino-disk-shaped sample with contact radii of 400 to 600 µm (Device V) were measured first with and then without a grounded top gate. The difference between charge pumping signals of both rings manifests the charge accumulation. The blue and red dashed lines represent the theoretical value of pumped charge at outer and inner ring separately. The orange dashed line represents the theoretical value of pumped charge without charge accumulation for both inner and outer rings (see Supplementary Materials for details). 
  (a-b) The charge pumping signals of the sample with (a) and without (b) a top gate at $f = 7.7977$ Hz. In-phase charge pumping data of part $\nu = 1$ plateaus are shown. Sketches of sample and measurement wires are represented in each figure. 
  (c) $f$ dependence of the in-phase inner-ring pumped charge at $\nu = 1$ is compared for both gate cases. Both signals are stable to $f$ variation.
  (d) In-phase charge pumping signals measured at each exact integer filling factors ($\nu_0$) are compared. The signals are divided by $\nu_0$ to have the same theoretical values for different $\nu_0$.}
  \label{fig:fig4}
\end{figure} 

The QHE superfluid with a gate chooses charge-mediated screening over the conventional mechanism of generating circulating currents to expel applied fields, governed by energy cost. With a nearby grounded gate providing large capacitance, the electrostatic energy required to draw in charge is substantially lower than that for direct current-mediated screening. Therefore, charge-mediated screening is expected to transform into current-mediated screening if the top gate is removed and the energy cost of charge accumulation becomes prominent.

Indeed, as shown in Fig. \ref{fig:fig4}(a-b), when the top gate is removed, charge accumulation almost vanishes in the plateaus, and nearly the same amount of pumped current is observed at both inner and outer contact rings. The underlying physics without a top gate is still CB superfluidity, but in an effective canonical ensemble, as opposed to the grand canonical ensemble with a top gate. For very small field variations, the system expels all flux that deviates from the exact filling—a Meissner effect—resulting in no charge accumulation. For larger fields, part of the flux penetrates as vortices, locally breaking CB superfluidity (i.e., the QHE) with electron density fixed within the vortex core, as it has already turned normal. Outside the vortex cores, the system remains in CB superfluidity with fixed electron density, so the total system exhibits no net charge accumulation. If the fields become even larger, CB superfluidity breaks down completely, entering another plateau region. CB superfluid's responses to external magnetic fields are similar to those of type-II superconductors, while the charge-mediated screening is an exclusive response giving lowest energy when the Coulomb potential increase is suppressed. Defining the Coulomb energy increase of charge-mediated screening with and without gate $E_1$ and $E_2$, energy cost of vortex-mediated screening $E_3$, and energy gain of CB condensate $E_0$, the ordering of these energies is ($E_1<E_3<E_2,\, E_0$) within quantized plateaus.

We also note that the observation of fixed electron density at $\nu=1$ excludes the scenario of band-structure QHE (single electron picture, as opposed to CB superfluidity, which can be understood purely from values of $\sigma_{xx}$ and $\sigma_{xy}$). The $\sigma_{xx}$ of our GaAs sample is small enough (below $1 \times 10^{-13}$ $\Omega^{-1}$ \cite{wang2026}), so that fully quantized charge accumulation would occur even without a top gate at the measurement frequency $f = 7.797$ Hz if the underlying mechanism were band-structure QHE (see Supplementary Materials for details). Furthermore, the quantized and vanished charge accumulation in both gate cases is stable to higher $f$, which manifests as stable charge pumping signals (Fig. \ref{fig:fig4}(c)), consistent with our energy analysis, which is independent of $f$. In comparison, the band-structure QHE scenario with fixed $\sigma_{xx}$ would expand the value of $Q_\text{a}$ when $f$ increases, as higher frequencies suppress decay. We also demonstrate that the vanished or quantized charge accumulation in the two gate cases is stable for different integer fillings (Fig. \ref{fig:fig4}(d)). Furthermore, we verified that the response of charge accumulation to the existence of gate is consistent across different samples and under different AC magnetic field amplitudes (potential variations) (see Supplementary Materials for details).

\section{Discussion and conclusion}

As demonstrated, the GaAs QH system manifests CB superfluidity, with the electron-to-flux-quantum ratio locked. This means all flux quanta and electrons condense into CBs to form superfluidity, with no net flux or electrons remaining—an effective Meissner effect minimizing energy. At first glance, this seems to require a fixed charge density and complete expulsion of extra magnetic field $\Delta B$ due to the Meissner effect.

However, multi-top-gate devices, fabricated to measure charge accumulation in edge-included and bulk-confined regions spatially, surprisingly show a uniformly quantized distribution $\rho = \nu (e^2/h) \Delta B$ regardless of location, seemingly contradicting this expectation. A closer look reveals that this observation aligns perfectly with CB superfluidity in a grand canonical ensemble, which is realistic given the contact electrodes. The uniformly quantized charge accumulation reflects the fixed electron-to-flux-quantum ratio, indicating electron transfer from contact electrodes to accommodate flux variations, preserving the Meissner effect by absorbing all free flux quanta. This process incurs far lower energy cost than direct magnetic field screening.

Here, the top gate plays an essential role in enabling the grand canonical ensemble, as it greatly reduces the Coulomb potential energy cost of extra electron density, allowing the energy gain from forming CBs to dominate. Indeed, when the top gate is removed, the system effectively switches to a canonical ensemble, maintaining fixed charge density under $\Delta B$. We have successfully tuned the system between effectively canonical and grand canonical ensembles by controlling the top gate, observing the accompanying changes in the Meissner effect and associated charge density.

This manuscript investigates the macroscopic response of CB superfluidity—charge accumulation involving macroscopic numbers of electrons and flux quanta, with applied magnetic fields up to $\sim 0.3$ Gs. When the applied field is greatly reduced so that the magnetic flux penetrating the entire sample area is $\lesssim \Phi_0 = h/e$, interesting responses are expected. Gradually increasing the applied field would eventually cause a single flux quantum $\Phi_0 = h/e$ to be suddenly dragged into the system, creating a pulsed current whose integration corresponds to one electron charge. Such single-flux experiments would reveal even richer physics involving the deep nature of composite particles consisting of electrons and flux quanta, inviting future exploration.

\backmatter

\section{Acknowledgements}
We acknowledge the support for the project by the National Key Research and Development Program of China (Grants No. 2021YFA1401600 and No. 2021YFA1401900) and the National Natural Science Foundation of China (Grant No. 12574184). The Princeton University portion of this project is funded in part by the Gordon and Betty Moore Foundation's EPiQS Initiative, Grant No. GBMF9615.01 to Loren Pfeiffer. We thank Mengqiao Geng, Liang Guo, Wenlu Lin, Xianghong Jin, Daiqiang Huang, and Zijian Zhou for discussions and experimental assistance. Y.Li thanks Lewei Li for the help with schematic preparation.

\section{Author contribution}
T.L. and Y.Liu conceived, designed, and supervised the project and coordinated collaborations among the research groups. Y.Li, R.W., and J.C. designed the experimental setup and performed the charge accumulation measurements. R.W. and Y.Li fabricated the GaAs quantum well wafers into devices. The wafers A and B were grown by W.Z. under the supervision of R.R.D. The wafers C were grown by A.G. and K.W.B. under the supervision of L.N.P. Y.Li, J.C., R.W., and T.L. analyzed the experimental data. The manuscript was written by Y.Li and T.L. with input from all authors. All authors discussed the results and provided feedback on the manuscript.

\section{Competing interests}
The authors declare no competing interests.

\bigskip

\bibliography{GaAs_superfluidity}

\begin{thebibliography}{10}
\expandafter\ifx\csname url\endcsname\relax
  \def\url#1{\burl{#1}}\fi
\expandafter\ifx\csname urlprefix\endcsname\relax\def\urlprefix{URL }\fi
\providecommand{\bibinfo}[2]{#2}
\providecommand{\eprint}[2][]{\url{#2}}
\providecommand{\doi}[1]{\url{https://doi.org/#1}}
\bibcommenthead

\bibitem{klitzing1980}
\bibinfo{author}{Klitzing, K.~v.}, \bibinfo{author}{Dorda, G.} \& \bibinfo{author}{Pepper, M.}
\newblock \bibinfo{title}{New method for high-accuracy determination of the fine-structure constant based on quantized hall resistance}.
\newblock \emph{\bibinfo{journal}{Physical review letters}} \textbf{\bibinfo{volume}{45}}, \bibinfo{pages}{494} (\bibinfo{year}{1980}).

\bibitem{tsui1982FQHE}
\bibinfo{author}{Tsui, D.~C.}, \bibinfo{author}{Stormer, H.~L.} \& \bibinfo{author}{Gossard, A.~C.}
\newblock \bibinfo{title}{Two-dimensional magnetotransport in the extreme quantum limit}.
\newblock \emph{\bibinfo{journal}{Physical Review Letters}} \textbf{\bibinfo{volume}{48}}, \bibinfo{pages}{1559} (\bibinfo{year}{1982}).

\bibitem{laughlin1983wavefunc}
\bibinfo{author}{Laughlin, R.~B.}
\newblock \bibinfo{title}{Anomalous quantum hall effect: an incompressible quantum fluid with fractionally charged excitations}.
\newblock \emph{\bibinfo{journal}{Physical Review Letters}} \textbf{\bibinfo{volume}{50}}, \bibinfo{pages}{1395} (\bibinfo{year}{1983}).

\bibitem{girvin1987}
\bibinfo{author}{Girvin, S.} \& \bibinfo{author}{MacDonald, A.~H.}
\newblock \bibinfo{title}{Off-diagonal long-range order, oblique confinement, and the fractional quantum hall effect}.
\newblock \emph{\bibinfo{journal}{Physical review letters}} \textbf{\bibinfo{volume}{58}}, \bibinfo{pages}{1252} (\bibinfo{year}{1987}).

\bibitem{read1989}
\bibinfo{author}{Read, N.}
\newblock \bibinfo{title}{Order parameter and ginzburg-landau theory for the fractional quantum hall effect}.
\newblock \emph{\bibinfo{journal}{Physical Review Letters}} \textbf{\bibinfo{volume}{62}}, \bibinfo{pages}{86} (\bibinfo{year}{1989}).

\bibitem{zhang1992}
\bibinfo{author}{Zhang, S.~C.}
\newblock \bibinfo{title}{The chern--simons--landau--ginzburg theory of the fractional quantum hall effect}.
\newblock \emph{\bibinfo{journal}{International Journal of Modern Physics B}} \textbf{\bibinfo{volume}{6}}, \bibinfo{pages}{25--58} (\bibinfo{year}{1992}).

\bibitem{haldane2011}
\bibinfo{author}{Haldane, F.}
\newblock \bibinfo{title}{Geometrical description of the fractional quantum hall effect}.
\newblock \emph{\bibinfo{journal}{Physical review letters}} \textbf{\bibinfo{volume}{107}}, \bibinfo{pages}{116801} (\bibinfo{year}{2011}).

\bibitem{gorkov1958}
\bibinfo{author}{Gor'Kov, L.}
\newblock \bibinfo{title}{On the energy spectrum of superconductors}.
\newblock \emph{\bibinfo{journal}{Sov. Phys. JETP}} \textbf{\bibinfo{volume}{7}}, \bibinfo{pages}{158} (\bibinfo{year}{1958}).

\bibitem{yang1962ODLRO}
\bibinfo{author}{Yang, C.~N.}
\newblock \bibinfo{title}{Concept of off-diagonal long-range order and the quantum phases of liquid he and of superconductors}.
\newblock \emph{\bibinfo{journal}{Reviews of Modern Physics}} \textbf{\bibinfo{volume}{34}}, \bibinfo{pages}{694} (\bibinfo{year}{1962}).

\bibitem{laughlin1981}
\bibinfo{author}{Laughlin, R.~B.}
\newblock \bibinfo{title}{Quantized hall conductivity in two dimensions}.
\newblock \emph{\bibinfo{journal}{Physical Review B}} \textbf{\bibinfo{volume}{23}}, \bibinfo{pages}{5632} (\bibinfo{year}{1981}).

\bibitem{1992_Russia_GaAs_chargepumping_and_Streda}
\bibinfo{author}{Dolgopolov, V.}, \bibinfo{author}{Shashkin, A.}, \bibinfo{author}{Zhitenev, N.}, \bibinfo{author}{Dorozhkin, S.} \& \bibinfo{author}{Von~Klitzing, K.}
\newblock \bibinfo{title}{Quantum hall effect in the absence of edge currents}.
\newblock \emph{\bibinfo{journal}{Physical Review B}} \textbf{\bibinfo{volume}{46}}, \bibinfo{pages}{12560} (\bibinfo{year}{1992}).

\bibitem{1995_GaAs_chargepumping}
\bibinfo{author}{Jeanneret, B.} \emph{et~al.}
\newblock \bibinfo{title}{Observation of the integer quantum hall effect by magnetic coupling to a corbino ring}.
\newblock \emph{\bibinfo{journal}{Physical Review B}} \textbf{\bibinfo{volume}{51}}, \bibinfo{pages}{9752} (\bibinfo{year}{1995}).

\bibitem{wang2026}
\bibinfo{author}{Wang, R.} \emph{et~al.}
\newblock \bibinfo{title}{Laughlin pumping assisted by surface acoustic waves}.
\newblock \emph{\bibinfo{journal}{arXiv preprint arXiv:2601.11921}}  (\bibinfo{year}{2026}).

\end{thebibliography}

\end{document}